\documentclass[12pt]{article}
\usepackage{amsfonts}
\title{Low-energy next-to-leading contributions to the effective action
in ${\cal N}=4$ SYM theory \footnote{Talk given by at the
International Workshop ``Supersymmetries and Quantum
 Symmetries'' (SQS 03), BLTP, JINR, Dubna, Russia, July 24-29, 2003.}}

\author{A.T. Banin\footnote{atb@math.nsc.ru},
N.G. Pletnev\footnote{pletnev@math.nsc.ru}}

\date{{\it
Institute of Mathematics, Novosibirsk, \\ 630090, Russia}}

\begin{document}

\maketitle

\begin{abstract}
Using formulation of ${\cal N}=4$ SYM theory in terms of ${\cal
N}=1$ superfields superfields we construct the derivative
expansion of the one-loop ${\cal N}=4$ SYM effective action in
background fields corresponding to constant Abelian strength
$F_{mn}$ and constant hypermultiplet. Any term of the effective
action derivative expansion can be rewritten in terms of ${\cal
N}=2$ superfields. The action is manifestly ${\cal N}=2$
supersymmetric but on-shell hidden ${\cal N}=2$ supersymmetry is
violated. We propose a procedure which allows to restore the
hidden ${\cal N}=2$ invariance.
\end{abstract}
\thispagestyle{empty}

\newcommand{\be}{\begin{equation}}
\newcommand{\ee}{\end{equation}}
\newcommand{\bea}{\begin{eqnarray}}
\newcommand{\eea}{\end{eqnarray}}

\section{Introduction}

The exact low-energy quantum dynamics of ${\cal N}=4$ SYM theory
in ${\cal N}=2$ vector multiplet sector is mastered by the
non-holomorphic effective potential ${\cal H}({\cal W},{\bar{\cal
W}})=\frac{1}{(4\pi)^2} \ln{\cal W}\ln\bar{\cal W}$, depending on
${\cal N}=2$ strengths ${\cal W}, {\bar{\cal W}}$ (see Refs.
\cite{6, bbku}). This result can be obtained entirely on the
symmetry grounds from the requirements of scale independence and
$R$-invariance up to a numerical factor \cite{6, nonr} as well as
by direct quantum field theory calculations (see e.g. \cite{bkts})
using various formulation of the model (${\cal N}=1$ superspace,
${\cal N}=2$ harmonic and ${\cal N}=2$ projective superspaces).

Recently, the complete exact low-energy effective action
containing the dependence both on ${\cal N}=2$ gauge superfields
and hypermultiplets has been discovered \cite{31}. The additional
hypermultiplet-dependent contributions containing the on-shell
${\cal W}, {\bar{\cal W}}$ and the hypermultiplet $q^{ia}$
superfields have been obtained on a purely algebraic ground and in
the harmonic supergraph calculations in the complete on-shell
${\cal N}=4$ supersymmetric form
\begin{equation}\label{2}
{\cal L}_{q}=c\left\{(X-1)\frac{\ln (1-X)}{X}+[{\rm
Li}_{2}(X)-1]\right\}, \quad X=-\frac{q^{ia}q_{ia}}{{\cal
W}\bar{\cal W}} ,
\end{equation}
where ${\rm Li}_{2}(X)$ is the Euler dilogarithm function and $c$
is a constant (see the details and denotations in Refs.
\cite{31}).

In the present work we briefly discuss the problem of derivation
of the subleading terms in the effective action, depending on all
fields of ${\cal N}=4$ supermultiplet and representation of these
terms in complete ${\cal N}=4$ supersymmetric form. This allows to
construct the derivative expansion of the one-loop effective
Lagrangian ${\cal L}_{eff}$ depending both on ${\cal N}=2$ gauge
background superfields, their spinor derivatives up to some order
and hypermultiplet background superfields.

\section{The background field method}
The simplest ${\cal N}=1$ background, which allows fulfilling
one-loop calculations, belongs to the Cartan subalgebra of the
gauge group $SU(2)$ spontaneously broken down to $U(1)$ and
constant space-time background hypermultiplet $q^{ia}$
\begin{equation}\label{3}
{\cal W}|=\Phi={\rm const}, \quad D^{i}_{\alpha}{\cal
W}|=\lambda^{i}_{\alpha}={\rm const},\quad q^{ia}|={\rm const},
\end{equation}
$$
D^{i}_{(\alpha}D_{\beta )i}{\cal W}|=F_{\alpha\beta}={\rm
const},\quad D^{\alpha (i}D^{j)}_{\alpha}{\cal W}|=0,\quad
D_{\alpha}^i q^{aj}|=0, \quad D_{\dot{\alpha}}^i q^{aj}|=0,
$$
where ${\cal W}, q^{ia}$ are ${\cal N}=2$ superfields carrying
on-shell components of the ${\cal N}=4$ vector multiplet.

The action of ${\cal N}=4$ SYM model is formulated in terms of
${\cal N}=1$ superspace as follows \cite{24}
\begin{eqnarray}
S &=& {1\over g^{2}}{\rm tr}\{\int d^{4}x
d^{2}\theta\,W^{2}+ \int d^{4}x
d^{4}\theta\,\bar{\Phi}_{i}{\rm e}^{V}\Phi^{i}{\rm
e}^{-V}+\nonumber\\
&+& {1\over 3!}\int d^{4}x d^{2}\theta\,
ic_{ijk}\Phi^{i}[\Phi^{j},\Phi^{k}]+{1\over 3!}\int
d^{4}x d^{2}\bar\theta\,
ic^{ijk}\bar\Phi_{i}[\bar\Phi_{j},\bar\Phi_{k}] \}.\label{classic}
\end{eqnarray}
All superfields here are taken in the adjoint representation of
the gauge group. In addition to the manifest ${\cal N}=1$
supersymmetry and $SU(3)$ symmetry on the $i,j,k,\ldots$ indices
of $\Phi$ and $\bar\Phi$, the action has the hidden global
supersymmetries (see e.g. \cite{15}).

We define one-loop  effective action $\Gamma$ depending on the
background superfields (\ref{3}) by a path integral over quantum
fields in the standard form
\begin{equation}\label{4}
{\rm e}^{i\Gamma} = \int \,{\cal D}v\,\,{\cal D}\varphi\, {\cal
D}c\,{\cal D}c'\, {\cal D}\bar{c}\,{\cal D}\bar{c}'\, {\rm
e}^{i(S_{(2)}+S_{\rm FP})},
\end{equation}
where $S_{(2)}$ is a quadratic in quantum fields part of the
classical action including a gauge-fixing condition and $S_{\rm
FP}$ is a corresponding ghost action. The main technical tool for
the ${\cal N}=1$-superfield calculations is the background
covariant gauge-fixing $ S_{\rm GF}=-{1\over g^{2}}\int
d^8z\,(F^{A}\bar{ F}^{A})$, with the convenient conditions for the
quantum superfields $v$ and ${\varphi}$
\begin{equation}
\bar{F}^{A}=\nabla^2v^{A} + [{1\over
\Box_{+}}\nabla^{2}\varphi^{i},\bar{\Phi}_{i}]^{A}~,\quad
F^{A}=\bar{\nabla}^2v^{A} + [ {1\over
\Box_{-}}\bar{\nabla}^2\bar{\varphi}_{i},\Phi^{i}]^{A}~,\label{f}
\end{equation}
where $\Box_+$, $\Box_-$ are the standard notations for
Laplace-like operators in the ${\cal N}=1$ superspace. The gauge
fixing functions (\ref{f}) can be considered as a superfield form
of so-called $R_{\xi}$-gauges (see Refs. \cite{23}). It should be
noted that the used gauge fixing doesn't preserve the hidden
${\cal N}=2$ supersymmetries. In gauge theories not all hidden
symmetries of the classical action can be maintained manifestly in
the quantization procedure, e.g.: on-shell supersymmetry,
(super)conformal symmetry (see current status of problem in
Ref.\cite{29}). According to the analysis given in Ref.\cite{30}
the problem of keeping the rigid symmetries manifest at the
quantum level is essentially equivalent to finding covariant gauge
conditions. In the case of conformal symmetry as well as hidden
supersymmetries of ${\cal N}=4$ SYM theory in ${\cal N}=1$ or
${\cal N}=2$ harmonic superspace such gauge conditions do not
exist. As a sequence, any rigid transformations has to be
accompanied by a field-dependent non-local gauge transformation in
order to restore the gauge slice.

\section{One-loop effective action expansion.}
The whole one-loop contribution to the effective action (\ref{4})
has an extremely simple form and is determined only by vector loop
contribution
\begin{equation}\label{O}
\Gamma = i\sum_{I<J}\,{\rm Tr}\ln (O_{V}-M)_{IJ}~,
\end{equation}
because ghost and hypermultiplet contributions mutually cancel
each other. The details of calculation are given in Ref.
\cite{15}. Such a functional trace has been already calculated by
different ways for models with one chiral background superfield
(see \cite{15, bkts} and reference therein). The difference
between the theory with and without hypermultiplets consists in
the replacement the value ${M}= \bar{\Phi}\Phi$ with the
$R$-symmetry group invariant
${M}=(\bar{\Phi}\Phi+\bar{Q}Q+\bar{\tilde{Q}}\tilde{Q})$.

The trace (\ref{O}) can be written as a power expansion of
dimensionless combinations $\Psi$, $\bar \Psi$ in vector and
hypermultiplet superfields, where
\begin{equation}
{\bar \Psi}^2 = \frac{1}{M^2} \,\nabla^2 {W^2}, \quad \Psi^2 =
\frac{1}{M^2} \,{\bar \nabla}^2 {\bar W}^2.
\end{equation}
In the constant field approximation this expansion can be summed
to the following expression for the whole one-loop effective
action (see details in \cite{bkts}):
\begin{equation}\label{n4gamma}
\Gamma = \frac{1}{8\pi^{2}}\int{\rm d}^8 z \int_{0}^{\infty}{\rm
d}t\,t\,{\rm e}^{-t}\frac{ W^{2}\bar{W}^{2}}{M^2}\,\omega (t\Psi,
t\bar{\Psi}),
\end{equation}
As a result, we see that the only difference between the effective
actions with and without the hypermultiplet background is
stipulated by $M=(\phi\bar\phi+f^{ia}f_{ia})$, where $\phi,
\bar\phi$ and $f^{ia}$ are physical bosonic fields of the ${\cal
N}=2$ vector multiplet and hypermultiplet. In component form, the
closed relation for one-loop effective action (\ref{n4gamma}) has
natural Schwinger-type expansion over $F^2/M^2$ powers which
doesn't include $F^6$ term that is a property of ${\cal N}=4$ SYM
theory \cite{bkts, Fr}. The expansion of the function $\omega$
defined in (\ref{n4gamma}) (see \cite{bkts}) induces the effective
action (\ref{n4gamma}) expansion in powers of $\Psi^2$,
$\bar{\Psi}^2$ as follows
\begin{equation}\label{gdecom}
\Gamma = \Gamma_{(0)}+\Gamma_{(2)}+\Gamma_{(3)}+\cdots,
\end{equation}
where the term $\Gamma_{(n)}$ in the bosonic sector corresponds to
$\Gamma_{(n)}\sim F^{4+2n}/M^{2+2n}$ and contains terms
$c_{m,l}\Psi^{2m}\bar{\Psi}^{2l}$ with $m + l = n$.

As it was shown in Ref. \cite{bkts, 15} any
$\Gamma_{(n)}$ term can be reconstructed to ${\cal N}=2$ form. In
particular, one can obtain an expression
\begin{equation}\label{g0}
\Gamma_{(0)}={1\over (4\pi)^{2}}\int d^{12}z \,\left(\ln {\cal
W}\ln \bar{\cal W}+ \sum_{k=1}^{\infty}{1\over k^{2}(k+1)}
X^{k}\right),
\end{equation}
where $X$ was defined in (\ref{2}). The second term in (\ref{g0})
can be transformed to the form (\ref{2}). We see that the
expression (\ref{g0}) is just the effective Lagrangian (\ref{2})
found in \cite{33}. Direct analysis also leads to the following
expression for $\Gamma_{(2)}$ ($\sim F^8$) in (\ref{gdecom})
\begin{eqnarray}\label{g2}
\Gamma_{(2)}&=& {1\over 2\cdot5!\cdot(4\pi)^{2}}\int d^{12}z\,
{\bf\Psi^{2}\bar{\Psi}^{2}} \left({1\over (1-X)^{2}}+{4\over (1-X)}\right.+\\
&+&\left.{6X-4\over X^{3}}\ln (1-X)+4{X-1\over
X^{2}}\right)~,\nonumber
\end{eqnarray}
where ${\bf\Psi}^{2}= {1\over {\cal W}^2} \bar{D}^4\ln \bar{\cal
W}$ is ${\cal N}=2$ scalar. Unfortunately, we can not guarantee
that the reconstructed effective action will be invariant under
the undeformed hidden ${\cal N}=$ supersymmetry.

\section{Construction of proper ${\cal N}=4$ supersymmetric effective action}

All obtained ${\cal N}=2$ supersymmetric contributions should not
be invariant under the undeformed hidden ${\cal N}=2$
supersymmetry transformation because of the background choice
(\ref{3}) and the gauge-fixing procedure (\ref{f}). The proper
${\cal N}=4$ calculations should take into account vector
derivatives along with hypermultiplet derivatives. It is obvious
that in order to obtain ${\cal N}=4$ supersymmetric contributions
from the ones given in the previous section, we have to add to
each term in the derivative expansion of (\ref{n4gamma}) some
extra terms containing fields $\lambda = W|$ of the vector
multiplet, which are presented in the effective action
(\ref{n4gamma}), as well as fields $\psi = D q|$ of the
hypermultiplet, which are absent in our calculations because of
the used background.

Let consider the on-shell ${\cal N}=4$ supersymmetric effective
action which is described by manifestly ${\cal N}=2$
supersymmetric effective Lagrangian depending on ${\cal W},
\bar{\cal W}$, their spinor derivatives, $q^{+}$ and spinor
derivatives of $q^{+}$. The superfield effective Lagrangian have
to be dimensionless and chargeless. The dimensional quantities
$D^{-}q^{+}, (D^{-}q^{+})^{2}, ...$ can be compensated by ${\cal
N}=2$ strengths and their spinor derivatives. Hence, any
contribution $\Gamma_{(n)}$ to the effective Lagrangian must be a
finite order polynomial in derivatives $D^{-}q^{+}$ with the
dimensionless coefficients $g_{n, k}(X)$ and some polynomial $P
_{n, k}(D^{l}{\cal W}, D^{l'}\bar{\cal W}, {\cal W}, \bar{\cal
W})$. Symbolically it can be written as follows
\begin{equation}\label{49}
\Gamma_{(n)} = \Gamma_{(n; 0)} + \Gamma_{(n; 1)} + \cdots +
\Gamma_{(n; k)}~, \; \Gamma_{(n; m)} = g_{n, m}(X)P_{n, m}(D^{l}{\cal
W}, D^{l'}\bar{\cal W}, {\cal W}, \bar{\cal W})(D^{-}q^{+})^{m}~,
\end{equation}
where $m$ corresponds to the power of the derivatives $D^{-}q^{+}$
and $\bar{D}^{-}q^{+}$ . If it is possible for some fixed $n$, the
polynomials (\ref{49}) under undeformed hidden ${\cal N}=2$
transformation should transform as
\begin{equation}\label{trans_rule}
\Gamma_{(n; m)} \rightarrow \Gamma_{(n; m)} \oplus \Gamma_{(n;
m+1)}~.
\end{equation}

The analysis of invariance is greatly simplified when we calculate
only the first ($D^+q^{-}$-independent) term in the expansion
(\ref{49}). One can examine that the obtained by direct
calculation part of $\Gamma_{(2)}$ term is  non-invariant under
hidden ${\cal N}=2$ transformation. The transformation structure
(\ref{trans_rule}) allows to construct ${\cal N}=4$ supersymmetric
on-shell term $\Gamma_{(2; 0)}$ because the transformation of the
other terms $\Gamma_{(2; k)}, k>0$ doesn't affect on it.

Let's suppose that $\Gamma_{(2; 0)}$ can be rewritten as
\begin{equation}
\Gamma_{(2; 0)} ={1\over 2(4\pi)^{2}}\int d^{12}z\, I~, \;\;
I=\sum_{n=0}^{\infty}I_{n}=\sum_{n=0}^{\infty}c_{n}{\bf
\Psi}^{2}{\bf \bar\Psi}^{2}\left({-2q^{a +}q^{-}_{a}\over {\cal
W}\bar{\cal W}}\right)^{n}.\label{complem_1}
\end{equation}
We rewrite all in terms of ${\cal N}=2$ harmonic superspace and
trace transformations evoked by parameters $\varepsilon^{\alpha
a}$. Saving only terms which can make contribution to the term
$\Gamma_{(2; 0)}$ one can find the variation of the general term
\begin{equation}\label{rest_gen}
\delta I_{n}= \delta^{(w)}I_{n} + \delta^{(q)}I_{n}=
I_{n}\left[-{(n+2)(n+6)\over(n+4)}{\delta \bar{\cal W}\over
\bar{\cal W}}\right]  + c_{n}{\bf \Psi}^{2}{\bf
\bar\Psi}^{2}\left({-2q^{b +}q^{-}_{b}\over {\cal W}\bar{\cal
W}}\right)^{n-1}\left[-4n {q^{a +}\delta q^{-}_{a}\over {\cal
W}\bar{\cal W}}\right]~.
\end{equation}
The chain of cancellation between variations $\delta^{(w)}I_{n+1}$
and $\delta^{(q)}I_{n}$ will occur when the recursion condition is
satisfied
\begin{equation}
c_{n}=c_{n-1}{(n+1)(n+5)\over n(n+3)}~\Rightarrow \,c_{n}={1\over
6\cdot5!}(n+5)(n+4)(n+1)~.\label{rec1}
\end{equation}

Summing the series (\ref{complem_1}), we find the proper leading
part $\Gamma_{(2; 0)}$ in the expansion (\ref{49}) of on-shell
${\cal N}=4$ supersymmetric $F^8$-term in the closed form
\begin{equation}\label{59}
\Gamma_{(2; 0)}={1\over 72}{1\over (4\pi)^{2}}\int d^{12}z
du\,{\bf \Psi}^{2}{\bf
\bar\Psi}^{2}{1-X+{3\over10}X^2\over(1-X)^4}.
\end{equation}
Thus, the leading bosonic part of complete on-shell ${\cal N}=4$
supersymmetric extension of $F^8$ invariant is finally
established.

\section{Summary}
We have studied the one-loop effective action in ${\cal N}=4$ SYM
theory, depending on ${\cal N}=2$ vector multiplet and
hypermultiplet fields. The calculations of superfield functional
determinants are done on specific ${\cal N}=1$ superfield
background corresponding to constant Abelian strength $F_{mn}$ and
constant hypermultiplet fields. The effective action depending on
all fields of ${\cal N}=4$ vector multiplet is restored on the
base of calculations only in quantum ${\cal N}=1$ vector multiplet
sector by functional arguments replacement (see (\ref{O}) and
(\ref{n4gamma})). Obtained results are presented in a manifest
${\cal N}=2$ supersymmetric form. The complete ${\cal N}=4$
supersymmetric low-energy effective action, which has been
discovered in \cite{31}, has been obtained. All terms (except the
leading one) in derivative expansion of the effective action are
not invariant under hidden ${\cal N}=2$ supersymmetry
transformations. We have considered the first subleading term in
expansion of the effective action in ${\cal N}=2$ vector multiplet
sector ($F^{8}$-term written via ${\cal N}=2$ superconformal
invariants depending on strengths ${\cal W}, \bar{\cal W}$ and
their spinor derivatives \cite{bkts}) and proved that it can be
completed up to on-shell ${\cal N}=4$ supersymmetric form by the
hypermultiplet dependent terms and presented as polynomial in
hypermultiplet spinor derivatives. The first leading term of this
polynomial, which depends on hypermultiplet but does not depend on
its derivatives, is given in explicit form (\ref{59}).

\section{Acknowledgements}
The authors thank the Organizing Committee of the SQS'03
conference for hearty welcome and partial support. The work was
supported in part by INTAS grant, INTAS-00-00254 and RFBR grant,
project No 03-02-16193. The authors also would like to thank
I.L.~Buchbinder for numerous discussion and the participation at
this work.


\begin{thebibliography}{000}



\bibitem{6} M. Dine, N. Seiberg, Phys.Lett. {\bf B409} (1997)
239;  D.A. Lowe, R. von Unge, JHEP {\bf 9811} (1998) 014;
 F. Gonzalez-Rey, M. Ro\v{c}ek, Phys.Lett. {\bf
B434} (1998) 303998; Periwal, R. von Unge, Phys.Lett. {\bf B430}
(1998) 71.

\bibitem{bbku} I.L. Buchbinder, S.M. Kuzenko, Mod.Phys.Lett. {\bf A13} (1998)
1623;  E.L. Buchbinder, I.L. Buchbinder and S.M. Kuzenko,
Phys.Lett. {\bf B446} (1999) 216; E.I. Buchbinder, I.L.
Buchbinder, E.A. Ivanov, S.M. Kuzenko, B.A. Ovrut, Physics of
Particles and Nuclei, {\bf 32} (2001) 641.

\bibitem {nonr}I.L. Buchbinder, S.M. Kuzenko, B.A. Ovrut, Phys.Lett. {\bf
B433} (1998) 335.


\bibitem{bkts} I.L. Buchbinder, S.M. Kuzenko, A.A. Tseytlin, Phys. Rev. {\bf
D62} (2000) 045001; I.L.~Buchbinder, A.Yu. Petrov, A.A. Tseytlin,
Nucl.Phys. {\bf B621} (2002) 179.

\bibitem{31} I.L. Buchbinder, E.A. Ivanov, Phys.Lett. {\bf
524} (2002) 208; "Exact ${\cal N}=4$ supersymmetric low-energy
effective action in ${\cal N}=4$ super Yang-Mills theory";
hep-th/0211067; to be published in the Proceedings of the 3-rd
Sakharov Conference on Physics, Moscow, May, 2002; I.L.
Buchbinder, E.A. Ivanov, A.Yu. Petrov, Nucl.Phys. {\bf B653}
(2003) 64.



\bibitem{24} J. Gates, M. Grisaru, M. Ro\v{c}ek and W. Siegel,
``Superspace", Benjamin/Cummings, Reading MA (1983).

\bibitem{15} N.G. Pletnev, A.T. Banin, Phys.Rev. {\bf D60} (1999) 105017;
A.T. Banin, I.L. Buchbinder, N.G. Pletnev, Nucl.Phys. {\bf B598}
(2001) 371; Phys.Rev. D {\bf 66}, 045021 (2002); Phys.Rev. D68
(2003) 065024.



\bibitem{23} B.A. Ovrut, J. Wess, Phys.Rev. D {\bf 25}, 409 (1982);
A.T. Banin, I.L. Buchbinder, N.G. Pletnev, Phys.Rev. D {\bf 66},
045021 (2002).


\bibitem{29} S.M. Kuzenko, I.N.
McArthur, Nucl.Phys. B{\bf 640} 78  (2002) and refernces therein.

\bibitem{30}J.W. van Holten, Phys.Lett. B{\bf 200} 507 (1988).

\bibitem{Fr}E.S. Fradkin and A.A. Tseytlin, Nucl.Phys. {\bf
B277} (1983) 252.

\bibitem{33}A. Galperin, E. Ivanov, S. Kalitzyn, V. Ogievetsky, E. Sokachev, {\it
Class. Quant. Grav.} {\bf 1}, 469 (1984); A.S. Galperin, E.A.
Ivanov, V.I. Ogievetsky, E.S. Sokachev, {\it Class. Quant. Grav.}
{\bf 2}, 601 (1985); {\bf 2}, 617 (1985); `` Harmonic
Superspace'', Cambridge Univ. Press, (2001).


\end{thebibliography}
\end{document}